\documentclass[12pt]{iopart}

\usepackage{graphicx}

 \newcommand\la{\langle}
 \newcommand\ra{\rangle}
 \newcommand\beq{\begin{equation}}
 
 \newcommand\eeq{\end{equation}}
 \newcommand\beqn{\begin{eqnarray}}
 \newcommand\eeqn{\end{eqnarray}}

\def\GeV{\,\mbox{GeV}}

\def\lsim{\mathrel{\rlap{\lower4pt\hbox{\hskip1pt$\sim$}}
    \raise1pt\hbox{$<$}}}         
\def\gsim{\mathrel{\rlap{\lower4pt\hbox{\hskip1pt$\sim$}}
 \raise1pt\hbox{$>$}}}         

\begin{document}

\title{Quenching of high-\boldmath$p_T$ hadrons: Alternative scenario}

\author{B.Z.~Kopeliovich$^{1-3}$, I.K.~Potashnikova$^{1}$ and
Ivan~Schmidt$^{1}$}

\address{$^1$Departamento de F\'{\i}sica y Centro de Estudios
Subat\'omicos,
Universidad T\'ecnica Federico Santa Mar\'{\i}a,
Casilla 110-V, 
Valpara\'\i so, Chile\\
$^2$ Institut f\"ur Theoretische Physik der Universit\"at,
Heidelberg, Germany\\
$^{3}$Joint Institute for Nuclear Research, Dubna, Russia}

 \begin{abstract}
 A new scenario, alternative to energy loss, for the observed suppression
of high-$p_T$ hadrons observed at RHIC is proposed. In the limit of a very
dense medium crated in nuclear collisions the mean free-path of the
produced (pre)hadron vanishes, and and the nuclear suppression, $R_{AA}$
is completely controlled by the production length. The RHIC data are well
explained in a parameter free way, and predictions for LHC are provided.
 \end{abstract}


The key assumption of the energy loss scenario for the observed
suppression of high-$p_T$ hadrons in nuclear collisions is a long
length of the quark hadronization which ends up the medium. This has
got no justification so far and was challenged in \cite{within}.

The quark fragmentation function (FF) was calculated in Born 
approximation in \cite{berger}: 
 \beq
\frac{\partial D^{Born}_{\pi/q}(z)}{\partial k^2}\propto
{1\over k^4}\,(1-z)^2\,,
\label{10}
 \eeq
 where $k$ and $z$ are the transverse and fractional longitudinal momenta
of the pion. One can rewrite this in terms of the coherence length
$l_c=z(1-z)E/k^2$, where $E$ is the jet energy. Then, $\partial
D^{Born}_{\pi/q}(z)/\partial l_c\propto (1-z)$, is $l_c$ independent.  
Inclusion of gluon radiation leads to the jet lag effect \cite{jetlag}
which brings $l_c$ dependence,
 \beq
\frac{\partial D_{\pi/q}(z)}{\partial l_c}\propto
(1-\tilde z)\,S(l_c,z)\ .
\label{20}
 \eeq
 Here $\tilde z=z[1+\Delta E(l_c)/E]$ accounts for the higher Fock
components of the quark, which are incorporated via the
vacuum energy loss $\Delta E(l_c)$ calculated perturbatively with a
running coupling. The induced energy loss playing a minor role is added as 
well. $S(l_c,z)$ is the Sudakov suppression caused by energy
conservation. Fig.~1 shows an example for the $l_c$-distributions
calculated for $z=0.7$ and different jet energies at $\sqrt{s}=200\GeV$.
 \begin{figure}[htbp]
\centerline{
  \scalebox{0.32}{\includegraphics{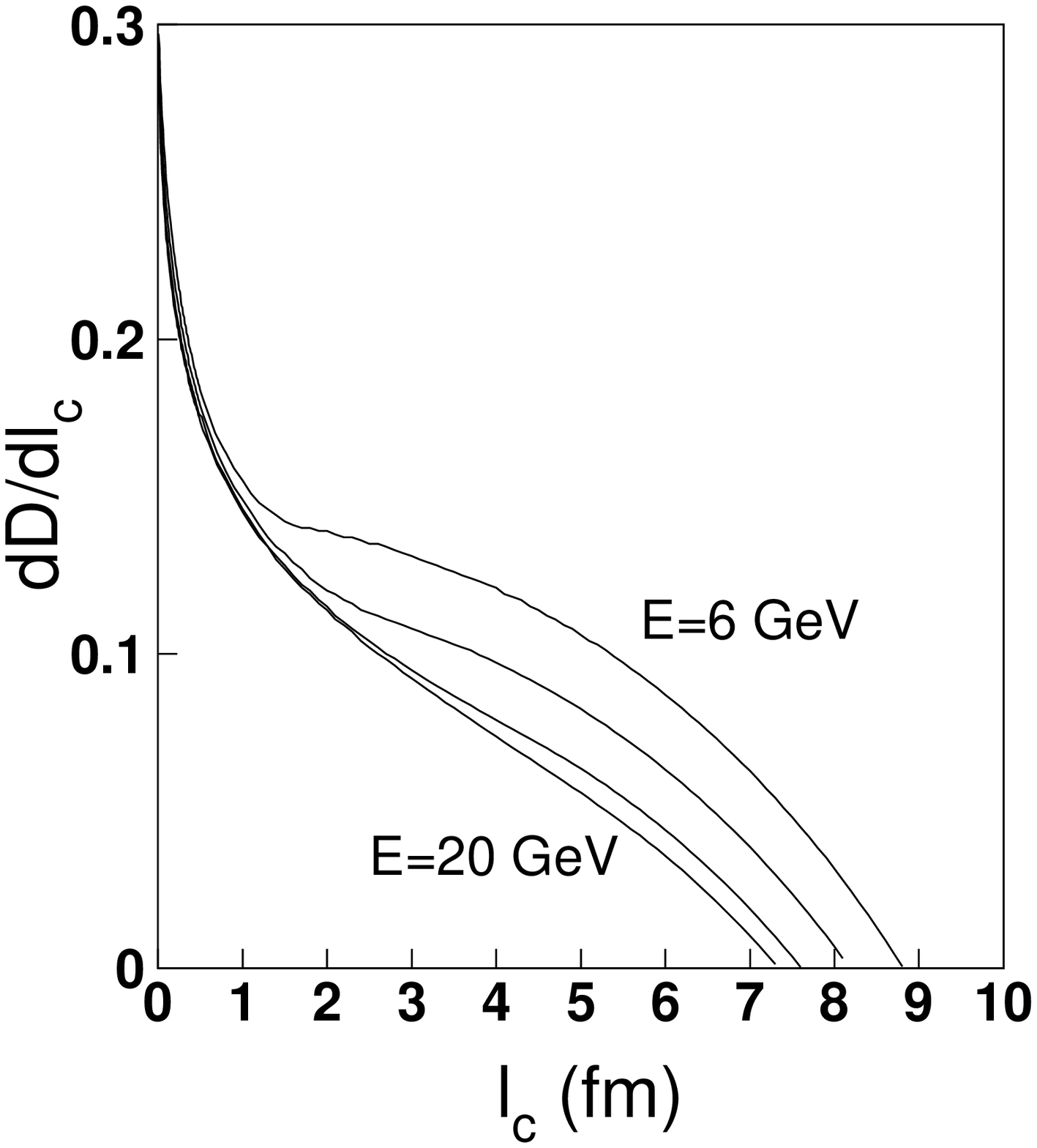}} 
  \scalebox{0.61}{\includegraphics{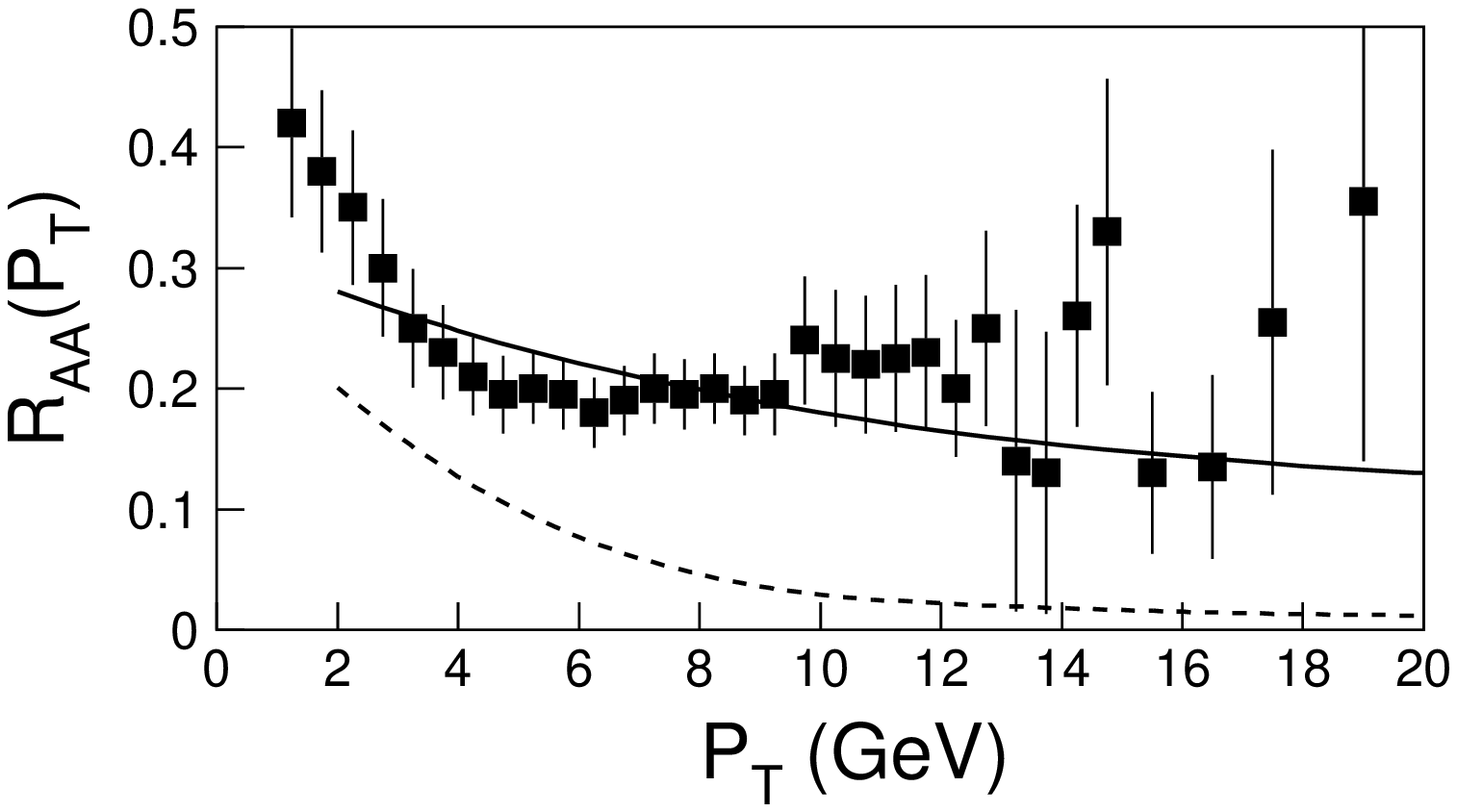}}
 }
 \caption{{\it Left:} $\partial D(z)/\partial l_c$
(in arbitrary units) at jet energies $6,\ 10,\ 16,\ 20\GeV$ and
$z=0.7$.  {\it Right:} Pion suppression in central $AA$ collisions  
($A\sim 200$) at $\sqrt{s}=200\GeV$ (solid) and $\sqrt{s}=5500\GeV$
(dashed). Data are from the PHENIX experiment.}
 \end{figure}

The pre-hadron, a $\bar qq$ dipole, may be produced with a rather large
initial separation $\la r_0^2\ra\approx 2l_c/E+1/E^2$ and it keeps
expanding.

To keep calculations analytic we consider a central, $b=0$, collision of
identical heavy nuclei with nuclear density
$\rho_A(r)=\rho_A\Theta(R_A-r)$.  Then we find,
 \beq
R_{AA}=\frac{\la l_c^2\ra}{R_A^2}\left[1-
A\,\frac{L}{\la l_c\ra} +
B\,\frac{L^2}{\la l_c^2\ra}
\right]\,,
\label{30}
 \eeq
 where the effective absorption length has the form,
$L^3=3p_T/(8\rho_A^2R_A\,X)$, and $X$ 
includes the unknown density of the medium and is to be
fitted to data on $R_{AA}$. However. if the medium is very dense,
i.e. $X$ is large, the last two terms in (\ref{30}) can be neglected, and
we can {\it predict} $R_{AA}$,
 \beq
R^h_{AA}=\frac{\la l_c^2\ra}{R_A^2}.
\label{40}
 \eeq
 With this expression we calculated $R_{AA}$ at the energies of RHIC and
LHC and in fig.~1 (right). This parameter free result well agrees with the
data supporting the assumption that the medium is very dense. Summarizing:
 \begin{itemize}
 \item
 The $A$-dependence, eq.~(\ref{40}), predicts $R_{AA}\approx0.42$
for $Cu-Cu$ confirmed by data.
 \item Vacuum radiation which
depends only on the current trajectory should be flavor independent. This
fact and the above consideration explains the strong suppression for heavy
flavors observed at RHIC.
 \item
 Since the strength of absorption does not affect $R_{AA}$, 
eq.~(\ref{40}), a single hadron and a pair of hadrons should be suppressed 
equally.
 \item The observed suppression $R_{AA}$ may not contain much information 
about the properties of the produced matter, it only says that the medium 
is very dense.

\end{itemize}

\ack
This work was supported in part by Fondecyt (Chile) grant 1050519 and
by DFG (Germany)  grant PI182/3-1.

\end{document}